\documentclass{article} % For LaTeX2e
\usepackage{iclr2027_conference,times}
\usepackage{graphicx} 
\usepackage{xspace}
\usepackage{amsmath,amsfonts,bm}

\def\eqref#1{equation~\ref{#1}}
\def\Eqref#1{Equation~\ref{#1}}
\def\1{\bm{1}}

\DeclareMathAlphabet{\mathsfit}{\encodingdefault}{\sfdefault}{m}{sl}
\SetMathAlphabet{\mathsfit}{bold}{\encodingdefault}{\sfdefault}{bx}{n}

\usepackage{hyperref}
\usepackage{url}
\usepackage{booktabs}       % professional-quality tables
\usepackage{wrapfig}

\newcommand{\method}{GEAR\xspace}

\title{Generative End-to-end Ad Retrieval \newline at Douyin}

\author{%
\textbf{Shaowen Zeng, Yanhua Huang, Jiacheng Sun, Jiarui Liu, Qian Dai, Zhikai Yang,} \\
\textbf{Hancheng Li, Boya Wu, Tuoyu Zhang, Yekui Chen, Xiang Sun} \\
ByteDance
}

\iclrfinalcopy % Uncomment for camera-ready version, but NOT for submission.

\begin{document}

\maketitle
\lhead{Preprint}

\begin{abstract}
Generative retrieval reformulates recommendation as the generation of discrete item tokens.
However, scaling this paradigm to real-world recommender systems reveals two critical bottlenecks:
1) \textbf{Representation collapse}, where the item tokenizer converges to degenerate results under \textit{continuous distribution shifts}, fundamentally hindering stable end-to-end adaptation.
2) \textbf{Item collisions}, where the \textit{massive candidate pool} causes distinct items to share identical token sequences, compromising the final retrieval precision.
Crucially, these bottlenecks are inherently coupled: expanding codebook capacity to mitigate collisions inevitably exacerbates collapse.
To address them simultaneously, we propose \textbf{\method}, an end-to-end framework that jointly optimizes the tokenizer, generator, and reranker.
To mitigate representation collapse, we introduce \textbf{BasisVQ}, which re-parameterizes the codebook via an orthogonal basis to enable global gradient sharing and rigid spatial rotation of the latent space, effectively stabilizing gradient dynamics without ad-hoc heuristics.
We further extend it to prefix-aware BasisRQ, substantially enhancing the codebook's expressiveness with the same asymptotic time complexity.
To resolve item collisions, \method integrates a context-conditioned reranking head into the generative process, efficiently disambiguating colliding items with minimal computational overhead. By unifying stable tokenization and joint reranking within an end-to-end generative framework, \method establishes a fully differentiable and scalable paradigm.
It currently serves hundreds of millions of daily active users on Douyin Ads, yielding substantial empirical improvements in extensive online A/B tests.
\end{abstract}

\section{Introduction}
Generative retrieval has demonstrated its success in a range of challenging scenarios~\citep{rajput2023recommender,zhou2025onerec,zhou2025onerecv2,wang2026pit,yang2025cobra}. 
In contrast to past paradigms, generative retrieval reformulates recommendation as a discrete item token generation task.
It employs an item tokenizer to map items into token sequences, followed by a token generator to autoregressively predict the subsequent item tokens.
However, when scaling this paradigm to real-world scenarios, this formulation introduces two critical bottlenecks.

The first bottleneck is \textbf{representation collapse}, well-known in computer vision~\citep{zhu2025addressing,mentzer2024fsq} but severely exacerbated in generative retrieval. Real-world recommenders are inherently dynamic systems, confronting \textit{continuous distribution shifts} as new items and user interests emerge. When updating the tokenizer via direct backpropagation on streaming data, this severe distribution shift exacerbates
gradient instability, trapping the codebook in underutilized sub-optimal states. To circumvent this,
previous works primarily rely on frozen tokenizers~\citep{zhou2025onerec, zhou2025onerecv2, xue2026generative}, or heuristic strategies~\citep{zheng2025pre,bai2025bi,yan2026merge} such as ad-hoc code re-initialization and exponential moving averages (EMA). However, frozen tokenizers inherently fail to adapt to continuous distribution shifts, and heuristics merely serve as weak surrogates for optimization
stability. Fundamentally, the lack of a principled scheme for stable, direct gradient optimization remains a critical roadblock for fully differentiable end-to-end learning.

The second bottleneck is \textbf{item collisions}, where the \textit{massive candidate pool} inherently causes distinct items to share
identical token sequences due to semantic similarity or limited codebook capacity. Conventional mitigation strategies defer disambiguation to a subsequent reranking stage~\citep{sun2026grank,zhou2024roger}, but introduce pipeline bottlenecks due to computational overhead or disjoint optimization. Recent works propose to append unique
tokens~\citep{rajput2023recommender,xue2026generative,yang2025liger}
to force distinction; however, this compromises the model's generalization ability, as unique tokens lack semantic structure.
More importantly, these two bottlenecks exhibit a
detrimental coupling: expanding the codebook capacity to alleviate collisions inevitably exacerbates representation collapse~\citep{zhu2025addressing,zhu2024scaling}. This intrinsic trade-off renders isolated solutions suboptimal, necessitating a unified framework to resolve them simultaneously.

Accordingly, we propose \textbf{\method}, a \textit{Generative End-to-end Ad Retrieval} framework featuring a training-serving co-design.
To mitigate representation collapse, we introduce \textbf{BasisVQ} to re-parameterize the codebook via a learnable orthogonal basis. Geometrically, optimizing this basis drives a rigid rotation of the entire coordinate system. This topology-preserving co-movement naturally recovers the utilization of inactive codes and effectively prevents vocabulary degeneration.
We further extend it to prefix-aware BasisRQ, providing a highly expressive vocabulary while maintaining the asymptotic time complexity. 
To resolve item collisions, \method integrates a lightweight reranking head that directly reuses the generator's autoregressive hidden states. Jointly optimized via a learning-to-rank objective, this context-conditioned head achieves fine-grained disambiguation among colliding items with minimal inference overhead.

By systematically resolving representation collapse and item collisions within a unified framework, \method provides a principled paradigm for fully differentiable
generative retrieval. We validate its robustness and scalability through extensive offline evaluations and large-scale online A/B tests on the Douyin Ads platform. 
Furthermore, supported by systemic serving optimizations, \method delivers efficient real-time serving at scale in production, supporting hundreds of millions of daily active users.

\section{Preliminaries}
% In this section, we first formalize the generative recommendation task. We then introduce
% item tokenization techniques, highlighting the underlying causes of codebook collapse and
% the inherent nature of item collisions.

Let $\mathcal{U}$ and $\mathcal{I}$ denote the sets of users and items, respectively. For
each user $u \in \mathcal{U}$, we utilize their historical behavior sequence $S_u = (i_1,
i_2, \dots, i_L)$ and denote the set of remaining features (e.g., age and
device type) as $X_u$. For each item $i \in \mathcal{I}$, we characterize it by a
comprehensive set of features $X_i$. Note that, in general, $X_i$ contains both static
features, such as multimodal information, and dynamic features, such as categorical
identifiers with learnable sparse embeddings. This hybrid feature space inherently
necessitates dynamic adaptation of the item representation during training.

\textbf{Generative retrieval} utilizes an item tokenizer to map the item
features $X_i$ to a sequence of $m$ ordered discrete tokens $T_i = (t_1, \dots, t_m)$.
The goal of the generative retrieval model is to autoregressively
generate the token sequence of the target
item~\citep{rajput2023recommender,zhou2025onerec}:
\begin{equation}
  P_{\mathcal{G}}(T_i | S_u, X_u) = \prod_{l=1}^{m} P_{\mathcal{G}}(t_l | S_u, X_u, t_0,
\dots, t_{l-1}),
\label{eq:regressive}
\end{equation}
where $\mathcal{G}$ denotes the autoregressive generator and $t_0$ is the begin-of-sequence (BOS) token.

\textbf{Item tokenization} typically follows the vector quantization (VQ)
paradigm~\citep{van2017neural,rajput2023recommender}.
Formally, given a continuous representation $z_i \in \mathbb{R}^{d}$ derived from the item
features $X_i$, we compress it into a discrete token $t_1 = \arg\min_k || z_i - c_{1,k} ||_2$, as the
index of the nearest code vector in the codebook $C_1 = \{c_{1,k}\}_{k=1}^{N_1}$,
where $N_1$ is known as the codebook size. To derive an ordered sequence of tokens, this
process is conventionally extended to residual quantization
(RQ)~\citep{xue2026generative,lee2022autoregressive} with a sequence of codebooks
$\{C_l\}_{l=1}^m$, where \(C_l \in \mathbb{R}^{N_l \times d}\) and \(N_l\) denotes the codebook size of the l-th quantization layer:
\begin{equation}
  t_l = \arg\min_k \left|\left| \left(z_i - \sum_{j=1}^{l-1} c_{j,t_j}\right) - c_{l,k}
\right|\right|_2.
\end{equation}

\section{\method}
\begin{figure}[ht]
    \centering
    \includegraphics[width=\linewidth]{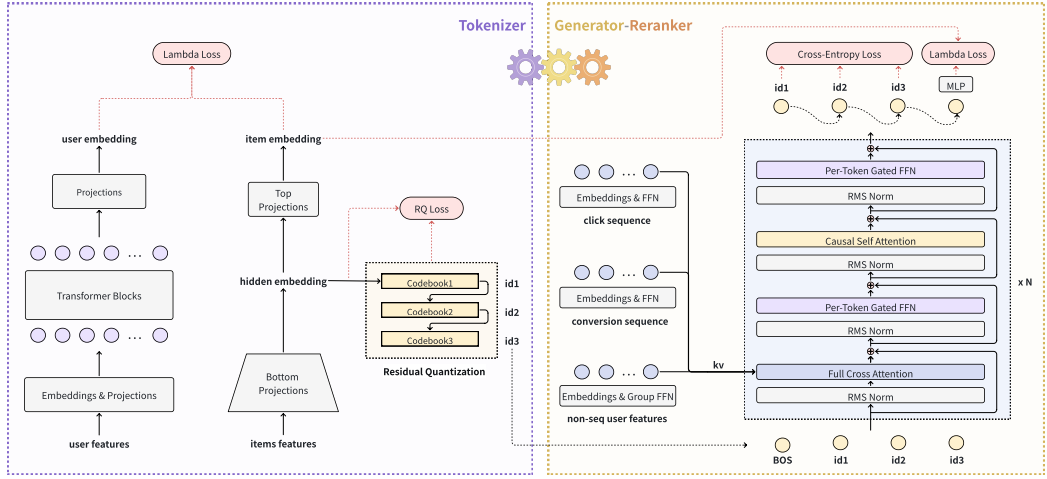}
    \caption{Overview of the \method framework. The item tokenizer (left) maps item features to discrete token sequences via residual quantization, and the generator-reranker (right) autoregressively decodes candidate sequences, with a lightweight reranking head that adds one extra decoding step to disambiguate colliding items. All components are jointly optimized end-to-end.}
    \label{fig:model_architecture}
\end{figure}

This section introduces \method (Figure~\ref{fig:model_architecture}), an end-to-end generative retrieval framework co-designed for training and serving. The following subsections describe its jointly optimized tokenizer, generator, and reranker, followed by training and serving details.

\subsection{Tokenizer} \label{sec:tokenizer}
In this subsection, we present our item tokenizer, which maps items in the recommendation corpus to discrete token sequences. We first introduce the derivation of dense item representations, then detail our proposed quantization scheme, prefix-aware BasisRQ—starting from the BasisVQ formulation with its convergence guarantee, geometric intuition, and complexity analysis, followed by its prefix-aware extension.

\textbf{Item Representation.} We adopt a two-tower architecture and extract the intermediate hidden embeddings from the item tower as dense representations, which are subsequently quantized by the prefix-aware BasisRQ module described below. As this two-tower design is standard practice in industrial retrieval and peripheral to our main contribution, we do not discuss it further.

\textbf{BasisVQ.} 
Let $C \in \mathbb{R}^{N \times d}$ denote the codebook in vanilla VQ.
In BasisVQ, we introduce a learnable basis matrix $W \in \mathbb{R}^{d \times d}$
to re-parameterize the codebook. To ensure orthogonality of the basis, we apply the Newton-Schulz iteration to approximately derive the orthogonal matrix $\hat{W} = h(W)$, where $h$
denotes the fully differentiable fixed-point iteration:
\begin{equation}
  \label{eq:newton-iter}
  W \leftarrow \frac{1}{2}W(3I - W^\top W).
\end{equation}
We then re-parameterize the codebook as $\hat{C} = C\hat{W}$, by which gradients will influence the entire codebook regardless of which specific code is selected. Furthermore, BasisVQ can
be naturally extended to residual quantization by applying it to all codebook layers, which we refer to as BasisRQ.

\textbf{Convergence Guarantee}. For a non-singular matrix, the Newton-Schulz iteration in \Eqref{eq:newton-iter} converges to an orthogonal matrix if and only if its spectral norm is strictly less than $\sqrt{3}$~\citep{higham1986computing}. To theoretically guarantee convergence and numerical stability during the forward pass, we normalize $W$ prior to the iteration:
\begin{equation}
W \leftarrow \frac{W}{\max\left(\|W\|_F, \epsilon\right)},
\end{equation}
where $\|W\|_F$ denotes the Frobenius norm and $\epsilon = 10^{-6}$ prevents division by zero. Since $\|W\|_2 \le \|W\|_F$, this step strictly bounds the initial spectral norm to $\|W\|_2 \le 1 < \sqrt{3}$.

\begin{figure}[ht]
    \centering
    \includegraphics[width=0.95\linewidth]{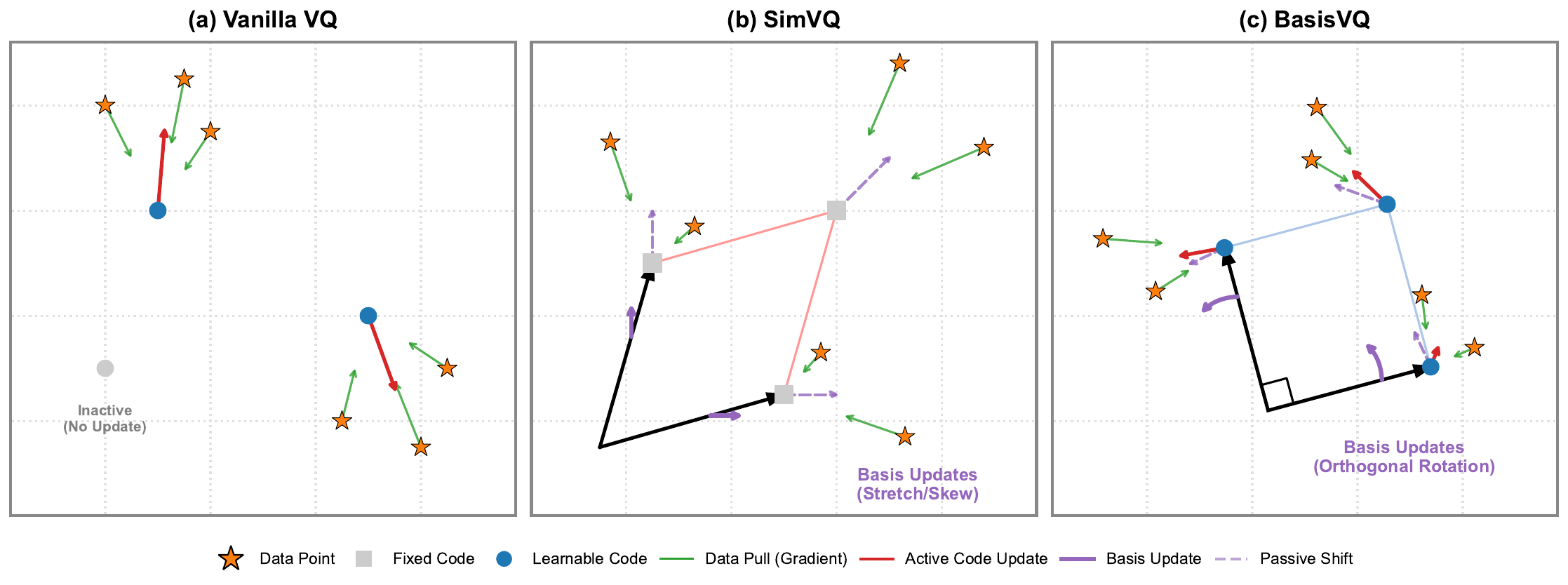}
    \caption{Comparison of VQ methods. Vanilla VQ suffers from sparse updates. SimVQ overlooks orthogonality and freezes the codebook to avoid collapse, limiting capacity. BasisVQ solves this via orthogonal bases, enabling simultaneous, fully dynamic updates for the entire codebook.}
    \label{fig:vq_methods}
\end{figure}

\textbf{Geometric Intuition.} While a linear projection $\hat{C} = CW$ allows gradient sharing, an
unconstrained basis $W$ is highly susceptible to collapse, as in SimVQ~\citep{zhu2025addressing}. We attribute this issue to the fact that inactive codes are often distorted or pushed
infinitely far from the data manifold, rendering them permanently unrecoverable.
BasisVQ resolves the issue by introducing orthogonality on bases. Geometrically, optimizing the basis via active codes induces a rigid rotation of the entire coordinate system. During this process, inactive codes are rotated
alongside active ones without distortion, strictly preserving their vector norms and pairwise spatial topology. Rather than remaining isolated, inactive codes
implicitly scan the latent space during the rotation. Particularly under the dynamic distribution shifts in large-scale streaming training, this topology-preserving co-movement increases the probability of inactive codes intersecting with emerging data manifolds, effectively stabilizing the overall gradient dynamics. Figure~\ref{fig:vq_methods} conceptually contrasts this mechanism with existing VQ methods. Furthermore, convergence experiments with synthetic data in Appendix~\ref{sec:toyexp}
demonstrate that BasisVQ effectively mitigates representation collapse compared to SimVQ.

\textbf{Complexity Analysis.} During training, in
each forward pass, it requires Newton-Schulz iterations to compute the
orthogonal basis $\hat{W}$. We find five iterations are empirically sufficient, which yields an overall computational
complexity of $\mathcal{O}(d^3)$, strictly bounded because of the independency of the batch size. In terms of space
complexity, BasisVQ introduces only a negligible $\mathcal{O}(d^2)$ parameter overhead for the basis
matrix $W$.
During inference, the re-parameterized codebook $\hat{C}$ can be pre-computed and cached
offline. Consequently, the
inference-time computational complexity remains strictly identical to that of vanilla VQ, achieving \textit{zero} latency overhead.

\textbf{Prefix-aware Extensions.} In conventional approaches (e.g., RQ-VAE and RQ-kmeans),
quantization layers are typically independent; for instance, the quantization in the second layer is
agnostic to the decision made in the first layer. This lack of prefix awareness often leads to sub-optimal codebook capacity. While recent work like QINCO~\citep{huijben2024residual} addresses this by
introducing an MLP transformation to dynamically adjust the codebook based on previous steps,
achieving significant improvements in reconstruction, its heavy computational overhead makes it inefficient for large-scale scenarios~\citep{vallaeys2025qinco2}. 

\begin{figure}[ht]
    \centering
    \includegraphics[width=0.75\linewidth]{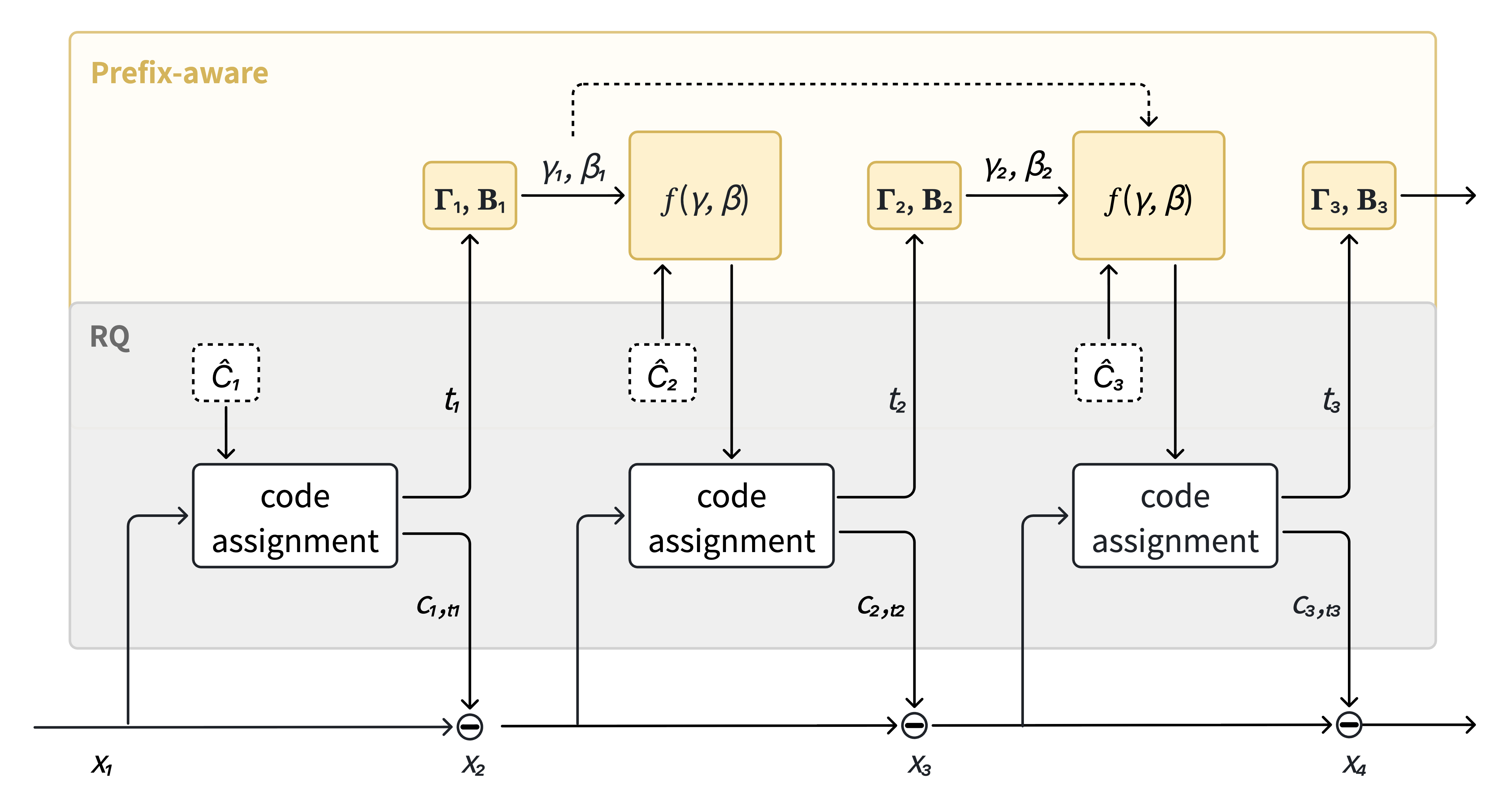}
    \caption{Prefix-aware residual quantization. The composite scaling \(\gamma\) and bias \(\beta\) are aggregated from all previous quantization decisions and applied as an affine transformation to the current layer's codebook.}
    \label{fig:contextual}
\end{figure}

We instead apply an element-wise affine
transformation to the codebook based on the quantization decisions of all previous layers, as illustrated in Figure~\ref{fig:contextual}. Empirically, we find that this achieves comparable performance with minimal computational overhead, thanks to a fast distance computation algorithm.
Formally, without loss of generality, consider the $l$-th quantization layer of BasisRQ. Let $x_l \in \mathbb{R}^{b
\times d}$ denote the input residual vectors in a batch. We retrieve sample-specific scaling
weights $\gamma_k \in \mathbb{R}^{b \times d}$ and biases $\beta_k \in \mathbb{R}^{b \times d}$ via a discrete lookup
from the per-layer parameter matrices $\Gamma_k, B_k \in \mathbb{R}^{N_k \times d}$ using the quantization codes
from each previous layer $k = 1, \dots, l-1$. Weights and biases from all previous layers are aggregated by iteratively computing the composite scaling $\gamma \in \mathbb{R}^{b \times d}$ and composite bias $\beta
\in \mathbb{R}^{b \times d}$:
\begin{equation}
\gamma = \bigodot_{k=1}^{l-1} \gamma_k, \quad \beta = \sum_{k=1}^{l-1} \left( \beta_k \odot \bigodot_{k^\prime=k+1}^{l-1} \gamma_{k^\prime} \right),
\label{eq:prefix-aware}
\end{equation}
where $\odot$ and $\bigodot$ denote element-wise multiplication and its generalized sequence product, respectively. Note that the above affine transformation is applied directly to the codebook rather than the input. However, a naive implementation of this sample-aware
codebook allocates a tensor of shape $(b, N_l, d)$, which leads to a severe memory overhead.
We thus define the bias-shifted residual as $\tilde{x}_l = x_l - \beta$ and algebraically rearrange the calculation of the distance matrix $D \in \mathbb{R}^{b \times N_l}$ by:
\begin{equation}
\label{eq:prefix-dist}
  D = \underbrace{\left(\tilde{x}_l \odot \tilde{x}_l\right) \mathbf{1}_d \mathbf{1}_{N_l}^\top}_{\text{Input Norms}} +
  \underbrace{\left(\gamma \odot \gamma\right) \left(\hat{C}_l \odot \hat{C}_l\right)^\top}_{\text{Scaled Codebook Norms}} -
  \underbrace{2 \left(\tilde{x}_l \odot \gamma\right) \hat{C}_l^\top}_{\text{Cross Term}},
\end{equation}
where $\mathbf{1}_N$ denotes an all-ones column vector of size $N$. Note that in vanilla VQ, computing the pairwise distance matrix also takes $\mathcal{O}(bN_ld)$ operations, i.e., our fast algorithm achieves the exact same asymptotic time complexity. We present a detailed proof procedure in the Appendix~\ref{sec:distance_derivation}.

\subsection{Generator}
The generator adopts an encoder-decoder architecture. The lightweight encoder projects raw user features into a compact key-value representation \(H\), while the decoder generates token sequences conditioned on \(H\), following the formulation in \Eqref{eq:regressive}.

\textbf{Encoder.} We extract two types of user features: non-sequence features (e.g.,
categorical attributes, statistical signals, and contextual information) and user
behavior sequences (without loss of generalization, we employ a click sequence and a conversion sequence). For non-sequence
features, we randomly partition their embeddings into $L_g$ groups and project each group via a dedicated Feed-Forward Network (FFN), yielding $H_g \in \mathbb{R}^{L_g \times d}$. For sequence features, the click sequence (length $L_c$) and the conversion sequence (length $L_e$) are each projected by an FFN, yielding $H_c \in \mathbb{R}^{L_c \times d}$ and $H_e \in \mathbb{R}^{L_e \times d}$.
The final KV representation, \( H = \mathrm{Concat}(H_g, H_c, H_e) \), serves as both the keys and
values in the cross-attention modules and is shared across all decoder
layers.

\textbf{Decoder.} The decoder is an \(N\)-layer Transformer that autoregressively generates the token sequence \((t_1, \dots, t_m)\). Each layer consists of cross-attention over \(H\), gated FFNs, causal self-attention, and RMSNorm applied before each sub-layer.

At step \(l\), let \(x \in \mathbb{R}^d\) be the input to the first layer, initialized as the embedding of the BOS token (\(t_0\)) and updated with the embedding of the previously generated token. Within each layer, \(x\) undergoes:
\begin{equation}
\begin{aligned}
x &= x + \mathrm{CrossAttn}(\mathrm{RMSNorm}(x), H), \\
x &= x + \mathrm{GatedFFN}(\mathrm{RMSNorm}(x)), \\
x &= x + \mathrm{CausalSelfAttn}(\mathrm{RMSNorm}(x)), \\
x &= x + \mathrm{GatedFFN}(\mathrm{RMSNorm}(x)),
\end{aligned}
\end{equation}
where \(\mathrm{GatedFFN}(x) = \mathrm{SiLU}(x W_1) \odot (x W_3) W_2\). The output of the final layer \(x_{\mathrm{out}} \in \mathbb{R}^d\) is projected to produce a probability distribution over the \(l\)-th vocabulary by:
\[
P(t_l \mid t_0, \dots, t_{l-1}, H) = \mathrm{Softmax}(x_{\mathrm{out}} W_{\mathrm{out}}^{(l)}),
\]
where \(W_{\mathrm{out}}^{(l)} \in \mathbb{R}^{d \times N_l}\) is the output projection.

\subsection{Reranker}\label{sec:reranker}
Although prefix-aware BasisRQ mitigates item collisions through a more expressive vocabulary, the \textit{massive candidate pool} inherently dictates that some items will still share the same token sequence and likelihood, rendering them indistinguishable to the generator. To enable fine-grained discrimination among such items according to user preferences, we augment our generative framework with a jointly learned reranking head for item-level scoring, as illustrated in Figure~\ref{fig:model_architecture}.

Specifically, after the decoder autoregressively generates the token sequence \((t_1, \dots, t_m)\), we perform one additional decoding step, whose output captures interactions among hierarchical item tokens and user features. We transform this output via an MLP into a user embedding \(u_{\mathrm{rerank}}\) and score each item as \(s_i^{\mathrm{rerank}} = u_{\mathrm{rerank}}^\top v_i\), where \(v_i\) is the item embedding produced by the tokenizer's item tower (Section~\ref{sec:tokenizer}).

This design enables fine-grained item discrimination with minimal computational overhead. It adds only one decoder step and lightweight dot-product scoring by reusing the encoder output \(H\) and item embeddings. 
Moreover, the reranker achieves precise ranking through a learning-to-rank objective. Since \(u_{\mathrm{rerank}}\) is conditioned on both the user and the generated token sequence, each candidate sequence yields its own user representation, dedicated to ranking the small set of colliding items under that sequence. This allows a simple dot product to achieve fine-grained discrimination comparable to heavier MLP-based interaction models.

\subsection{Training and Serving} \label{sec:training_serving}

We jointly train the tokenizer, generator, and reranker on our dataset. For each request, we sample a set of items with ground-truth labels \(y_i\), where \(y_i > y_j\) indicates item \(i\) is preferred to item \(j\).
To improve training stability, we adopt a progressive warm-up schedule that first trains the two-tower model in the tokenizer with \(\mathcal{L}_{\mathrm{LTR}}\) for an initial number of steps, then activates \(\mathcal{L}_{\mathrm{RQ}}\) for quantization, and finally activates \(\mathcal{L}_{\mathrm{Gen}}\) and \(\mathcal{L}_{\mathrm{Rerank}}\), preventing the generator from relying on an unstable tokenizer.

\textbf{Tokenizer Loss.} We train the two-tower model with LambdaLoss~\citep{wang2018lambdaloss}:
\begin{equation}
\mathcal{L}_{\mathrm{LTR}} = \sum_{i,j: y_i > y_j} \log\left(1 + e^{-\sigma(s_i - s_j)}\right) \cdot |\Delta \mathrm{NDCG}_{ij}|,
\label{eq:ltr}
\end{equation}
where \(s_i\) is the dot product of the two tower outputs, \(\sigma\) is a scaling factor, and \(\Delta \mathrm{NDCG}_{ij}\) is the change in NDCG from swapping items \(i\) and \(j\).

The quantization module is trained with codebook loss and commitment loss:
\begin{equation}
\mathcal{L}_{\mathrm{RQ}} = \sum_{l=1}^{m} \left( \left\| \mathrm{sg}[x_l] - c_l^* \right\|_2^2 + \eta \left\| x_l - \mathrm{sg}[c_l^*] \right\|_2^2 \right),
\label{eq:rq_loss}
\end{equation}
where \(x_l\) and \(c_l^*\) are the residual and matched code at layer \(l\), respectively, \(m\) is the number of quantization layers, \(\mathrm{sg}[\cdot]\) denotes the stop-gradient operator, and \(\eta\) balances the two terms.

\textbf{Generator Loss.} The generator is trained with teacher forcing: the target token sequence from the tokenizer is provided as input, with causal masking to prevent information leakage. To emphasize high-value samples, we adopt secpm reweighting, where secpm is a revenue-based metric from the ranking stage. The loss is the sum of weighted cross-entropy over the vocabulary across all steps:
% The loss is the weighted cross-entropy over the token vocabulary at each step:
\begin{equation}
\mathcal{L}_{\mathrm{Gen}} = - \sum_{l=1}^{m} w \cdot \log P(t_l \mid t_0, \dots, t_{l-1}, H),
\label{eq:gen}
\end{equation}
where \(w = \log(1 + \mathrm{secpm})\) for impressed samples and \(w = 0\) otherwise.

\textbf{Reranker Loss.} The reranker is trained with LambdaLoss:
\begin{equation}
\mathcal{L}_{\mathrm{Rerank}} = \sum_{i,j: y_i > y_j} \log\left(1 + e^{-\sigma(s_i^{\mathrm{rerank}} - s_j^{\mathrm{rerank}})}\right) \cdot |\Delta \mathrm{NDCG}_{ij}|,
\label{eq:rerank}
\end{equation}
where \(y_i\) and \(\Delta \mathrm{NDCG}_{ij}\) follow the same definitions as in \Eqref{eq:ltr}.

\textbf{Training Objective.} The overall training objective is a weighted combination of the four losses:
\begin{equation}
\mathcal{L} = \lambda_{\mathrm{LTR}} \mathcal{L}_{\mathrm{LTR}} + \lambda_{\mathrm{RQ}} \mathcal{L}_{\mathrm{RQ}} + \lambda_{\mathrm{Gen}} \mathcal{L}_{\mathrm{Gen}} + \lambda_{\mathrm{Rerank}} \mathcal{L}_{\mathrm{Rerank}},
\label{eq:joint}
\end{equation}
where the \(\lambda\) coefficients balance the contribution of each loss term.

\textbf{Inference.} During inference, the generator decodes autoregressively from the BOS token using beam search. At each step \(l\), we retain the top \(K_l\) token sequences, ultimately obtaining \(K_m\) candidate sequences. For each candidate sequence, we retrieve its associated items and apply the reranker to produce item-level scores for fine-grained disambiguation among colliding items. The system-level implementation of this inference pipeline is detailed in Section~\ref{sec:deployment}.

\section{Related Work}
% \subsection{Tokenization}
\textbf{Tokenizer.} Item tokenizer bridges continuous item representations and the discrete output space of generative models. VQ~\citep{van2017neural} is the dominant paradigm for this purpose, mapping the embeddings to the nearest codes in a learnable codebook.

A fundamental challenge in VQ is representation collapse: only a small fraction of codes receive gradient updates, leaving the majority ``dead'' and underutilized. One line of work freezes the tokenizer after pre-training on static data~\citep{zhou2025onerec, zhou2025onerecv2, xue2026generative}, which bypasses collapse but fails to adapt to dynamic distributions. Another employs heuristic updates such as EMA and ad-hoc re-initialization of inactive codes~\citep{zheng2025pre, bai2025bi, yan2026merge}, which improve stability but yield suboptimal results. Neither achieves dynamic adaptation through direct gradient optimization. More recently, SimVQ~\citep{zhu2025addressing} and GRIT-VQ~\citep{you2026gritvq} both apply a learnable linear transformation to the codebook, ensuring that gradients can reach all codes. SimVQ keeps the base codebook frozen and the transformation unconstrained, which can distort inactive codes and further impede their recovery. GRIT-VQ improves gradient stability with a generalized radius surrogate, at the cost of non-trivial complexity.

RQ~\citep{lee2022autoregressive} extends VQ with multi-layer codebooks for finer granularity, but treats each layer independently, limiting expressiveness. To endow RQ with prefix awareness, QINCo~\citep{huijben2024residual} introduces MLPs conditioned on previous quantization steps, at prohibitive computational cost; PSRQ~\citep{wang2025psrq} offers a lighter alternative with marginal gains.

In contrast, our prefix-aware BasisRQ mitigates representation collapse through an orthogonal transformation under streaming updates, while capturing prefix dependencies via lightweight affine transformations—without increasing the asymptotic time complexity of vanilla RQ.

\textbf{Generative Retrieval.} 
Generative retrieval reformulates recommendation as autoregressive sequence generation over discrete item tokens.
TIGER~\citep{rajput2023recommender} first established this paradigm with hierarchical semantic IDs learned through RQ and an encoder-decoder Transformer.
Subsequent work has focused on scaling this paradigm to real-world settings.
The OneRec series~\citep{zhou2025onerec,zhou2025onerecv2} replaces the conventional cascading pipeline with a generative architecture.
NEZHA~\citep{wang2026nezha} reduces decoding latency via lightweight prediction heads.
GR4AD~\citep{xue2026generative} and GPR~\citep{zhang2025gpr} further adapt the paradigm to advertising, with value-aligned optimization and hierarchical decoders respectively.

Recent studies have begun to revisit item tokenizer.
MTGRec~\citep{zheng2025pre} improves semantic representations through multi-token pre-training, yet still treats tokenization as an independent stage.
PIT~\citep{wang2026pit} establishes a feedback loop from generation to tokenization, but relies on heuristic search-and-select strategies rather than end-to-end optimization.
However, stable joint optimization of tokenization and generation remains largely unexplored.

Another line of work enhances item-level disambiguation.
Early approaches appended unique tokens to disambiguate colliding items~\citep{rajput2023recommender, xue2026generative}, at the cost of semantic generalization.
Gryphon~\citep{tikhonovich2026gryphon} tackles item collisions with a jointly trained item-level scoring module, but incurs non-trivial computational overhead.
LIGER~\citep{yang2025liger} and COBRA~\citep{yang2025cobra}, while not addressing collisions, combine generative retrieval with dense item representations for item-level scoring, yet neither is optimized for advertising goals nor supports lightweight deployment.

To address these issues, \method jointly optimizes the tokenizer, generator, and reranker, resolving item collisions with minimal computational overhead while enabling dynamic tokenizer adaptation.

\section{System Deployment}\label{sec:deployment}
\begin{figure}[ht]
    % \vspace{-3pt}
    \centering
    \includegraphics[width=0.95\linewidth]{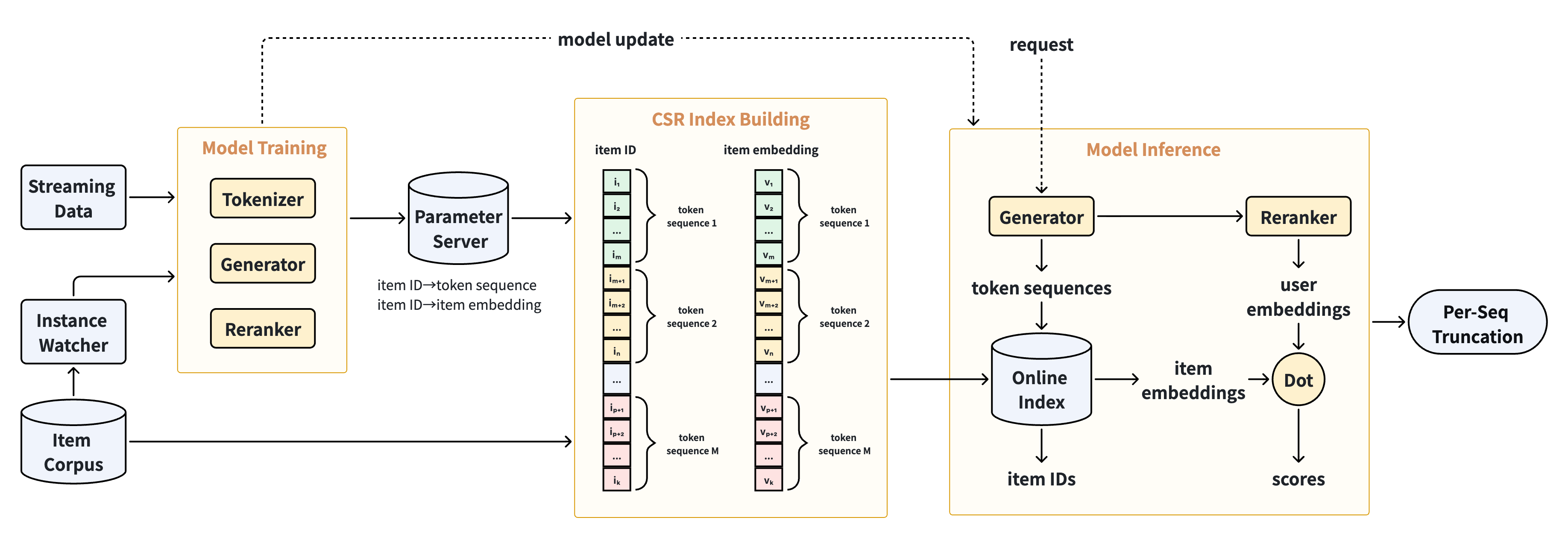}
    \vspace{-0.2em}
    \caption{Deployment architecture of \method, comprising streaming training, nearline indexing, and online inference.}
    \label{fig:deployment_architecture}
    \vspace{-0.7em}
\end{figure}

Figure~\ref{fig:deployment_architecture} illustrates the deployed architecture, comprising streaming training, nearline indexing, and online service. \method is deployed as an additional retrieval channel complementing the existing retrieval systems. All item-dependent computation is decoupled from the critical inference path, and the token sequence generation, sequence-to-item expansion, and item-level scoring are integrated into a unified computation graph. 
The system delivers efficient real-time serving in production, supporting hundreds of millions of daily active users.

\textbf{Streaming Training and Nearline Indexing.}
The Parameter Server (PS) maintains an entry for each item, consisting of its token sequence and item embedding. As the model is trained on streaming data, these entries are continuously updated. An instance watcher periodically refreshes existing items and incrementally encodes newly activated items; both operations require only a single forward pass through the item tower and tokenizer. A nearline index builder then collects all active entries and organizes them into a compressed sparse row (CSR)-like index, where a sorted directory maps each token sequence to two aligned lists of the corresponding items and their precomputed embeddings. The resulting index snapshot is refreshed at minute-level granularity and distributed to the online inference system along with the model artifact.

\textbf{Online Inference Pipeline.}
Given a user request, the generator performs KV-cached beam search to decode \(K_m\) candidate token sequences.
For these sequences, the system retrieves their associated items and precomputed embeddings from the online index. 
The reranker (Section~\ref{sec:reranker}) then computes sequence-specific user embeddings and scores the items via low-precision dot products, as both user and item embeddings are quantized to signed packed-INT4 format.
Finally, items within each sequence are sorted by rerank scores and truncated to a per-sequence limit, preventing any single token sequence from dominating the final results and promoting candidate diversity.

\section{Large-scale Real-world Experiments}
In this section, we report results from online A/B tests on the Douyin Ads platform, followed by ablation studies and a dedicated analysis of the tokenizer.
We adopt a three-layer prefix-aware BasisRQ tokenizer with codebook sizes \( (2048, 512, 128) \), and use beam widths \( (256, 256, 512) \) for generation. Full implementation details are provided in Appendix~\ref{sec:appendix_impl}.

\subsection{Overall Performance}
Comprehensive online A/B tests are conducted on the Douyin Ads platform over a 7-day period, with each treatment group allocated 5\% of the total traffic. We evaluate performance using two key revenue-oriented metrics: ADSS (Advertiser Score) and ADVV (Advertiser Value)~\citep{chai2025longer,jiang2026tokenmixer}.

\textbf{Main Results.} Our proposed \method is deployed as an additional generative retrieval pathway alongside existing dense retrieval systems. The baseline is the production system without this additional pathway; all other components remain identical. The online test yields consistent and significant improvements: ADSS increases by +0.563\%, and ADVV increases by +0.658\%. These gains, observed in a production system serving hundreds of millions of daily active users, demonstrate the practical value of our approach.

\textbf{Ablation Studies.} To isolate the contribution of each core component, we conduct ablation experiments on a subset of ad candidates. Table~\ref{tab:ablation} reports the relative improvements over the baseline for each variant. We consider three ablations: (i) replacing our tokenizer with vanilla RQ (\textbf{w/o prefix-aware BasisRQ}); (ii) removing the joint reranking head, with colliding items randomly truncated (\textbf{w/o rerank}); (iii) disabling the secpm reweighting in \(\mathcal{L}_{\mathrm{Gen}}\) (\textbf{w/o secpm reweight}). The largest drop is observed when removing prefix-aware BasisRQ, highlighting that a tokenizer that can dynamically adapt to distribution shifts while maintaining high expressiveness is crucial for effective generative retrieval. The reranking head also improves performance by resolving item collisions through fine-grained scoring, and secpm reweighting provides an additional gain.

\begin{table}[ht]
\vspace{-0.9em}
\centering
\small
\caption{Online A/B test results on the Douyin Ads platform in ADSS and ADVV: relative gains over the production baseline for subset-level ablation studies and full global deployment.}
\label{tab:ablation}
\begin{tabular}{lcc}
\toprule
\textbf{Method} & \textbf{ADSS (\%)} & \textbf{ADVV (\%)} \\
\midrule
\textit{Subset Evaluation} & & \\
\quad \method & \textbf{+0.446} & \textbf{+0.498} \\
\quad w/o prefix-aware BasisRQ & +0.288 & +0.219 \\
\quad w/o rerank & +0.358 & +0.341 \\
\quad w/o secpm reweight & +0.395 & +0.354 \\
\midrule
\textit{Global Deployment} & & \\
\quad \method & \textbf{+0.563} & \textbf{+0.658} \\
\bottomrule
\end{tabular}
\vspace{-0.3em}
\end{table}

\textbf{Cold-start Analysis.} Beyond the overall revenue gains, \method also benefits cold-start advertising by generating discrete token sequences that generalize better than item-ID-based embeddings for new ads without historical interaction data. We measure this via the cold-start activation rate, defined as the proportion of new advertisers reaching a predefined traffic threshold within a given time window. \method achieves a relative improvement of +1.46\% on this metric.

\subsection{Tokenizer Analysis}
\label{sec:tokenizer_analysis}
To assess the impact of tokenizer design choices in the full model, we compare the following configurations: (i)~prefix-aware BasisRQ of size \( (2048, 512, 128) \); (ii)~prefix-aware BasisRQ of size \( (512, 512, 512) \); (iii)~BasisRQ without prefix awareness of size \( (2048, 512, 128) \); (iv)~vanilla RQ with EMA updates of size \( (2048, 512, 128) \). We evaluate these variants using three metrics computed on a held-out evaluation set: (i)~Codebook Utilization (Util.), defined as the proportion of code sequences from all residual codebooks activated by at least one item; (ii)~Item Collision Rate (Coll.), defined as the proportion of items that share their code sequence with at least one other item, measuring item discriminability; (iii)~Quantization Error (MSE) between the original dense representation and its quantized reconstruction, measuring information retention. 

Table~\ref{tab:tokenizer} summarizes the results. Our proposed tokenizer achieves the highest codebook utilization, the lowest collision rate, and the lowest quantization error. Compared with the uniform schedule \( (512,512,512) \), the progressive configuration delivers superior performance under the same representational capacity. 
Removing prefix awareness increases collision rate (+11.5 p.p.) and quantization error (+0.21$\times 10^{-3}$), validating enhanced codebook expressiveness.
Finally, BasisRQ achieves substantially higher utilization (+3.23 p.p.) and lower collision rate (-20.4 p.p.) than vanilla RQ, confirming the effectiveness of gradient sharing via orthogonal bases.

\begin{table}[ht]
\vspace{-0.9em}
\centering
\small
\caption{Comparison of tokenizer configurations.}
\label{tab:tokenizer}
\begin{tabular}{lccc}
\toprule
\textbf{Method} & \textbf{Util. (\%)} & \textbf{Coll. (\%)} & \textbf{MSE (\(\times 10^{-3}\))} \\
\midrule
Prefix-aware BasisRQ (\(2048,512,128\)) & \textbf{8.83} & \textbf{52.3} & \textbf{1.82} \\
Prefix-aware BasisRQ (\(512,512,512\)) & 6.16 & 75.6 & 2.04 \\
BasisRQ (\(2048,512,128\)) & 7.45 & 63.8 & 2.03 \\
Vanilla RQ (\(2048,512,128\)) & 4.22 & 84.2 & 2.32 \\
\bottomrule
\end{tabular}
\vspace{-0.5em}
\end{table}

\section{Conclusion}
We present \method, an end-to-end generative retrieval framework that jointly optimizes the tokenizer, generator, and reranker to address representation collapse and item collisions in large-scale industrial recommenders. Prefix-aware BasisRQ mitigates representation collapse via orthogonal transformations and enhances expressiveness with prefix awareness, while a lightweight reranker further resolves item collisions. Online A/B tests on the Douyin Ads platform—serving hundreds of millions of daily active users—yield consistent revenue gains. Future work will explore more efficient decoding to further reduce inference latency and broaden deployment scenarios.

\section*{Generative AI Disclosure Statement}
During the preparation of this manuscript, we used generative AI tools for proofreading, sentence polishing, and assisting in checking mathematical equations for consistency and errors. The tool was not used to generate original research content or experimental data. The authors reviewed, verified, and accepted all edited text, taking full responsibility for the final manuscript.

\bibliography{reference}
\bibliographystyle{iclr2027_conference}

% \section{Appendix}
\clearpage
\appendix
\section{BasisVQ VS SimVQ on Sythetic Data}
\label{sec:toyexp}
We provide a toy experiment to simulate the distribution shift inherent in streaming training. Specifically, we generate a synthetic dataset in 8-dimensional space consisting of
$K=512$ widely dispersed Gaussian clusters, yielding a total of 128,000 data points. In real-world
streaming scenarios, as new items and user interests constantly emerge, the data distribution
dynamically shifts away from historical data. For these newly emerged distributions, the historical
codebook is typically located far away, a phenomenon mathematically equivalent to a highly suboptimal
initialization. To explicitly mimic this streaming challenge, we initialize all code vectors
within a small hypercube near the origin, drawn from a uniform distribution $[-0.1, 0.1]$.
Consequently, the initial codebook is located extremely far from the true data manifold. Under this
setting, the vast majority of codes receive zero point assignments during the initial forward passes,
instantly becoming ``dead codes''. Note that all evaluated methods strictly share identical
initialization method, optimizer, and loss function to ensure fairness.

\begin{figure}[h]
    \centering
    \includegraphics[width=\linewidth]{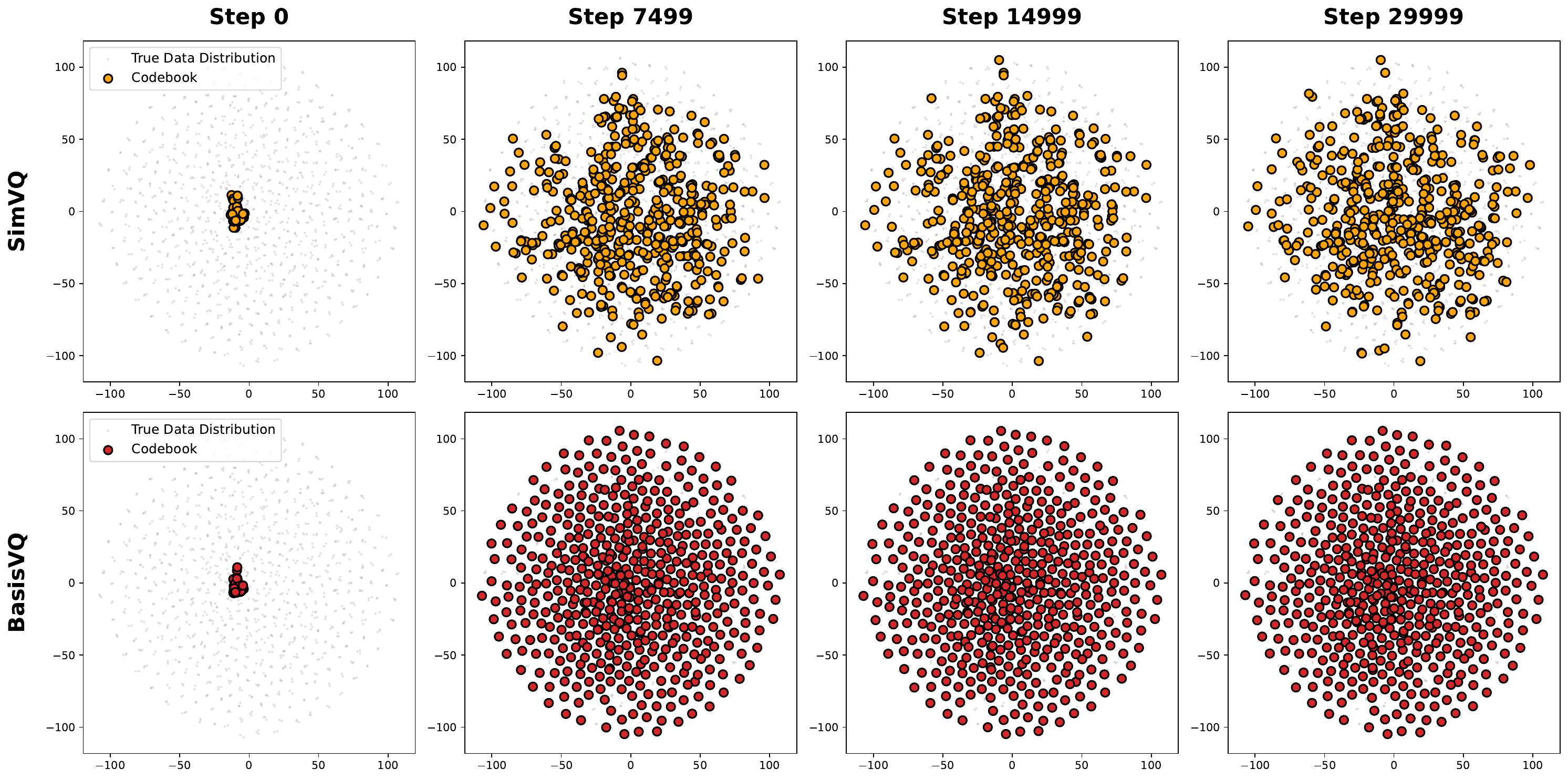}
    \caption{Visual comparison of SimVQ and BasisVQ on a challenging toy dataset. The
initial codebook is located far from the true data manifold, explicitly simulating real-world scenarios where the approximate data distribution is not known in advance.}
    \label{fig:pca_toy}
\end{figure}

To visualize the
spatial evolution of the codebooks, we project the data and codes onto a 2D plane
using t-SNE~\citep{van2008visualizing}. To prevent projection bias and ensure visual consistency, a
single global t-SNE model is fitted on the entire true data manifold, and all subplots share strictly
identical coordinate axes.

The visualization of the training procedure is shown in Figure~\ref{fig:pca_toy}. By utilizing a shared basis matrix, SimVQ
activates more codes but ultimately fails to explore the full data manifold because its codebook
parameters are frozen. In contrast, the proposed BasisVQ delivers the best performance. The rigid
rotation mechanism successfully resurrects the initial dead codes, allowing them to uniformly explore
and cover the entire data manifold with a higher rate of convergence.

\section{Detailed Derivation of the Fast Distance Computation}
\label{sec:distance_derivation}

In this section, we provide a step-by-step derivation for the L2-distance matrix $D$ introduced in \Eqref{eq:prefix-dist}. Let $x \in \mathbb{R}^{b \times d}$ be the input residual vectors, $\hat{C} \in \mathbb{R}^{N_m \times d}$ be
the base codebook, and $\gamma \in \mathbb{R}^{b \times d}$ and $\beta \in \mathbb{R}^{b \times d}$ be the prefix-
specific parameters (composite scaling and composite bias, respectively).

Conceptually, the transformed codebook for the $n$-th sample in a batch is given by $\tilde{C}_{(n)} = \gamma_{(n)}
\odot \hat{C} + \beta_{(n)}$. The L2 distance between the $n$-th input $x_{(n)}$ and the $k$-th transformed codeword is:
\begin{equation}
    D_{(n, k)} = \left\| x_{(n)} - (\gamma_{(n)} \odot \hat{C}_{(k)} + \beta_{(n)}) \right\|_2^2
\end{equation}

To avoid materializing the dynamically transformed codebook as a dense 3D tensor of shape $(b, N_m, d)$, we define the
bias-shifted residual vector as $\tilde{x}_{(n)} = x_{(n)} - \beta_{(n)}$. Rearranging the terms, we get:
\begin{equation}
    D_{(n, k)} = \left\| \tilde{x}_{(n)} - \gamma_{(n)} \odot \hat{C}_{(k)} \right\|_2^2
\end{equation}

Expanding the squared L2 norm along the feature dimension $d$:
\begin{equation}
\begin{aligned}
    D_{(n, k)} &= \sum_{j=1}^{d} \left( \tilde{x}_{(n,j)} - \gamma_{(n,j)} \hat{C}_{(k,j)} \right)^2 \\
    &= \underbrace{\sum_{j=1}^{d} \tilde{x}_{(n,j)}^2}_{\text{Term 1}} + \underbrace{\sum_{j=1}^{d}
\left(\gamma_{(n,j)}^2 \hat{C}_{(k,j)}^2\right)}_{\text{Term 2}} - \underbrace{2 \sum_{j=1}^{d} \left(\tilde{x}_{(n,j)}
\gamma_{(n,j)}\right) \hat{C}_{(k,j)}}_{\text{Term 3}}
\end{aligned}
\end{equation}

We can vectorize the computation for the entire batch and codebook to obtain the distance matrix $D \in \mathbb{R}^{b
\times N_m}$:
\begin{itemize}
    \item \textbf{Term 1 (Input Norms):} The sum of squares of $\tilde{x}$ can be computed as $(\tilde{x} \odot
\tilde{x}) \mathbf{1}_d$. To broadcast it across all $N_m$ codewords, we multiply by $\mathbf{1}_{N_m}^\top$, yielding
$(\tilde{x} \odot \tilde{x}) \mathbf{1}_d \mathbf{1}_{N_m}^\top$.
    \item \textbf{Term 2 (Scaled Codebook Norms):} The dot product of squared $\gamma$ and squared $\hat{C}$ is
naturally expressed as the matrix multiplication $(\gamma \odot \gamma) (\hat{C} \odot \hat{C})^\top$.
    \item \textbf{Term 3 (Cross Term):} The cross-correlation term is computed via the matrix multiplication $-2
(\tilde{x} \odot \gamma) \hat{C}^\top$.
\end{itemize}

Summing these vectorized terms yields the exact efficient formulation presented in the main text:
\begin{equation}
  D = \underbrace{\left(\tilde{x} \odot \tilde{x}\right) \mathbf{1}_d \mathbf{1}_{N_m}^\top}_{\text{Input Norms}} +
  \underbrace{\left(\gamma \odot \gamma\right) \left(\hat{C} \odot \hat{C}\right)^\top}_{\text{Scaled Codebook Norms}} -
  \underbrace{2 \left(\tilde{x} \odot \gamma\right) \hat{C}^\top}_{\text{Cross Term}}
\end{equation}

This algebraic rearrangement ensures that the bottleneck operations are standard matrix multiplications—$(\gamma \odot
\gamma) (\hat{C} \odot \hat{C})^\top$ and $(\tilde{x} \odot \gamma) \hat{C}^\top$—both of which execute in $\mathcal{O}(b
N_m d)$ time. Thus, it strictly matches the asymptotic time complexity of vanilla VQ without incurring
the prohibitive memory overhead of 3D tensors.

\section{Implementation Details}
\label{sec:appendix_impl}
\textbf{Tokenizer.} The output embeddings of the two-tower model are set to a dimensionality of 96. The quantization module is a three-layer Prefix-aware BasisRQ with codebook sizes \( (2048, 512, 128) \) and a code vector dimension of 128. This progressive decrease in capacity aligns with the diminishing residual energy across quantization layers, yielding superior performance under the same total codebook capacity.

\textbf{Generator.} For the encoder, non-sequential user features are grouped into \( L_g = 28 \) tokens; the click sequence length \( L_c = 200 \); the conversion sequence length \( L_e = 100 \). The decoder consists of 2 layers, each employing 6 attention heads with a head dimension of 64, and the Gated FFN has an intermediate dimension of 384.

\textbf{Reranker.} The user representation \( u_{\mathrm{rerank}} \) for reranking are set to dimension 96.

\textbf{Losses and Training.} The scaling factor \( \sigma \) is set to 1 for both LambdaLoss terms (\Eqref{eq:ltr} and \Eqref{eq:rerank}), and the RQ loss hyperparameter \( \eta \) (\Eqref{eq:rq_loss}) is 0.1. The loss weights \( \lambda_{\mathrm{LTR}}, \lambda_{\mathrm{RQ}}, \lambda_{\mathrm{Gen}}, \lambda_{\mathrm{Rerank}} \) are \( 1, 1, 10, 1 \) (\Eqref{eq:joint}), respectively. During inference, beam widths of 256, 256, and 512 are used for the three generation steps.

\end{document}